\documentstyle[12pt,epsf]{article}
\global\let\epsfloaded=Y

\begin{document}
\begin{titlepage}

\vspace{0.2in}
\rightline{\vbox{\halign{&#\hfil\cr
&NTUTH-97-04\cr
&July 1997\cr}}}

\vfill

\begin{center}
{\Large \bf  Perspectives on Quark Mass and Mixing Relations
\footnote
{
Talk presented at the Workshop on Masses and Mixings of 
Quarks and Leptons, University of Shizuoka, Japan, March 19 -- 21, 1997.
To appear in Proceedings.
}
}
\vfill
        {\bf George Wei-Shu HOU}
\footnote{
E-mail: wshou@phys.ntu.edu.tw.} \\

        {Department of Physics, National Taiwan University,}\\
        {Taipei, Taiwan 10764, R.O.C.}\\
\end{center}
\vfill
\begin{abstract}
Two down--up perspectives on quark mass--mixing relations are reviewed.
The modified Fritzsch path
relates $V_{cb}$ and $V_{ub}$ to $M_u$, 
but has trouble with the low $V_{cb} \simeq 0.04$ value.
The modified Wolfenstein path focuses on the change in
$V_{ub}$ from $\lambda^3 \mbox{\ (ca. 1983)} \leadsto \lambda^4 \mbox{\ (ca. 1994)}$.
The relations
$m_d/m_s\sim m_s/m_b \sim \delta \sim \lambda^2 
\simeq \vert V_{cb}\vert \simeq \vert V_{us}\vert^2$
and
$m_u/m_c\sim m_c/m_t \sim \delta^2 \sim \lambda^4 \sim \vert V_{ub}\vert$
suggest a clean separation of the origin of $V_{\rm KM}$:
$\vert V_{us}\vert \equiv \lambda$ and
$\vert V_{cb}\vert \equiv \delta$ arise from $M_d$ 
while $V_{ub} \equiv B\lambda^4 e^{-i\phi}$ comes from $M_u$.
Five to six parameters might suffice for ten mass--mixing parameters,
with $\delta$ seemingly 
the more sensible expansion parameter,
while $\lambda \simeq \sqrt{m_d/m_s}$ 
is tied empirically to  $(M_d)_{11} = 0$.
The approximate relations suggest a near weak scale origin of flavor.
\end{abstract}
\vfill
\end{titlepage}
\begin{center}

{\large\bf  Perspectives on Quark Mass and Mixing Relations}
\\[.3in]

{\bf  George Wei-Shu Hou}\\
Department of Physics\\
National Taiwan University\\
Taipei, Taiwan 10764, R.O.C.

\end{center}

\vglue.2in
\begin{abstract}
Two down--up perspectives on quark mass--mixing relations are reviewed.
The modified Fritzsch path
relates $V_{cb}$ and $V_{ub}$ to $M_u$, 
but has trouble with the low $V_{cb} \simeq 0.04$ value.
The modified Wolfenstein path focuses on the change in
$V_{ub}$ from $\lambda^3 \mbox{\ (ca. 1983)} \leadsto \lambda^4 \mbox{\ (ca. 1994)}$.
The relations
$m_d/m_s\sim m_s/m_b \sim \delta \sim \lambda^2 
\simeq \vert V_{cb}\vert \simeq \vert V_{us}\vert^2$
and
$m_u/m_c\sim m_c/m_t \sim \delta^2 \sim \lambda^4 \sim \vert V_{ub}\vert$
suggest a clean separation of the origin of $V_{\rm KM}$:
$\vert V_{us}\vert \equiv \lambda$ and
$\vert V_{cb}\vert \equiv \delta$ arise from $M_d$ 
while $V_{ub} \equiv B\lambda^4 e^{-i\phi}$ comes from $M_u$.
Five to six parameters might suffice for ten mass--mixing parameters,
with $\delta$ seemingly 
the more sensible expansion parameter,
while $\lambda \simeq \sqrt{m_d/m_s}$ 
is tied empirically to  $(M_d)_{11} = 0$.
The approximate relations suggest a near weak scale origin of flavor.
\end{abstract}

%
\def\ltap{\ \raisebox{-.5ex}{\rlap{$\sim$}} \raisebox{.4ex}{$<$}\ }
\def\gtap{\ \raisebox{-.5ex}{\rlap{$\sim$}} \raisebox{.4ex}{$>$}\ }
\pagestyle{plain}

\vskip0.5cm
\noindent{\large\bf I. Introduction}
\vskip0.3cm

Why am I slightly embarrassed in playing the mass-mixing game (MMG)?
Though MMG as started by \cite{Fritzsch} Harald Fritzsch is fascinating,
I always hesitated in writing something.
I mean,
{\it just how many papers can fit in a $3\times 3$ matrix?}

This attitude is not unique to me, as the depressing 
{\it philosophical} debates with
referees (when I fail the temptation\ldots) often show:
\begin{description}
\item[\ \ \, \bf --] ``Long on Words and Short on Substance. \ldots\ {\it ad hoc}."\\
 ------ Indeed we are!
\item[\ \ \, \bf --] ``No Real Insight!"\\
``Papers should contain More than Assumptions!"
\item[\ \ \, \bf --] ``Void of Dynamics!"\\
$\hookrightarrow$ ``Ans\" atze not based on Symmetry 
{\it do not go beyond {\sl\bf Mere Numerology!}"}\\
$\hookrightarrow$ ``There is an \boldmath{$\infty$} of phenomenologically consistent 
and predictive Ans\" atze\\
\phantom{xxx}  that Do Not Follow from Symmetry."\\
$\hookrightarrow$ ``{\bf I do Not think they should be Published.}"
\end{description}

Well, does this ring a bell? Perhaps you even secretly agree when 
not sitting in the author's pants?
I would like to make the following,
\vskip0.4cm
\centerline{STATEMENT}
\begin{verse}
\vskip -0.2cm Flavor question very difficult!\\
\  ----- Available info limited. Just ask Mendeleev, Balmer, or Gell-Mann.\\
But fermion masses and mixing dominate \# of free parameters in SM!\\
\  ----- Taking $m_\nu = 0$ and neglecting $\theta_{\rm QCD}$,
one has $13/18 > 1/2$.\\
\vskip 0.3cm
\hskip 1.5cm $\Longrightarrow$ \ \ \ {\large Reduction of Parameters Desirable!}
\end{verse}

\vskip -0.2cm Physics has always progressed on Idealization and 
establishing Empirical Rules or regularities, 
way before (and usually facilitating) Dynamical Explanations!
So, I shall proceed with good conscience, 
since the {\it True Origins} of $m_f$, $V_{\rm KM}$ 
and CP violation are still {\it Unknown}. 
And, afterall, it's the subject of this workshop.

\vskip 0.6cm

There has been three paths in the mass-mixing game 
(as compared to much more serious ``theory",
such as from topology, chaos, etc.):
\begin{enumerate}
\item Fritzsch \cite{Fritzsch} ($\sqrt{m_i/m_j}$): 
                   this will be our first theme \cite{HeHou}.
\item Wolfenstein \cite{Wolf} ($\lambda$-expansion): 
                   this will be our second, main, theme \cite{HW}.
\item Dimopoulos--Hall--Raby \cite{DHR} (texture): 
                  will be commented on briefly.
\end{enumerate}

Before we proceed, it is instructive to gain some perspective on the 
{\it scale} of $m_f$ generation.
We tend to think that ``we" are normal, and ``they" are abnormal,
hence the top is {\it very heavy},
as we are made of $u$, $d$ and $e$ and need $\nu_e$ to get energy.
Clearly $m_t \gg m_d$. 
Likewise, the $W$, $Z$, as well as $H^0$ bosons are also very heavy.
But we should remember that all masses in SM are 
pinned to the v.e.v. scale, $v \sim 250$ GeV.
We do not really know what happens at v.e.v. scale 
or beyond (call it ``heaven").
From this perspective, as illustrated in Fig. 1, 
in fact the ``heavy particles" appear to be {\it normal}, 
while we, you and I, seem to be made of
{\it zero modes}. 
It is the latter that needs to be explained ({\it e.g.} pions as Goldstones),
rather than the states with mass $\sim$ dynamical scale,
in analogy with QED and QCD.
This suggests that more states could appear 
at or above the v.e.v. scale.
In this light, the mass-mixing hierarchy
reflects  a restoration of Decoupling (in spontaneously broken 
nonabelian gauge theories):
Heavier Quarks do not play a major role in light quark physics
(low energy phenomena).
Fortified with this thought, let's play game:
Mass-Mixing relations without peeking above the v.e.v. scale.

\begin{figure*}[htb]
\let\picnaturalsize=N
\def\picsize{3.8in}
\def\picfilename{shizuoka1.eps}
\ifx\nopictures Y\else{\ifx\epsfloaded Y\else\input epsf \fi
\global\let\epsfloaded=Y
\vskip -2.9cm
\centerline{\ifx\picnaturalsize N\epsfxsize \picsize\fi \epsfbox{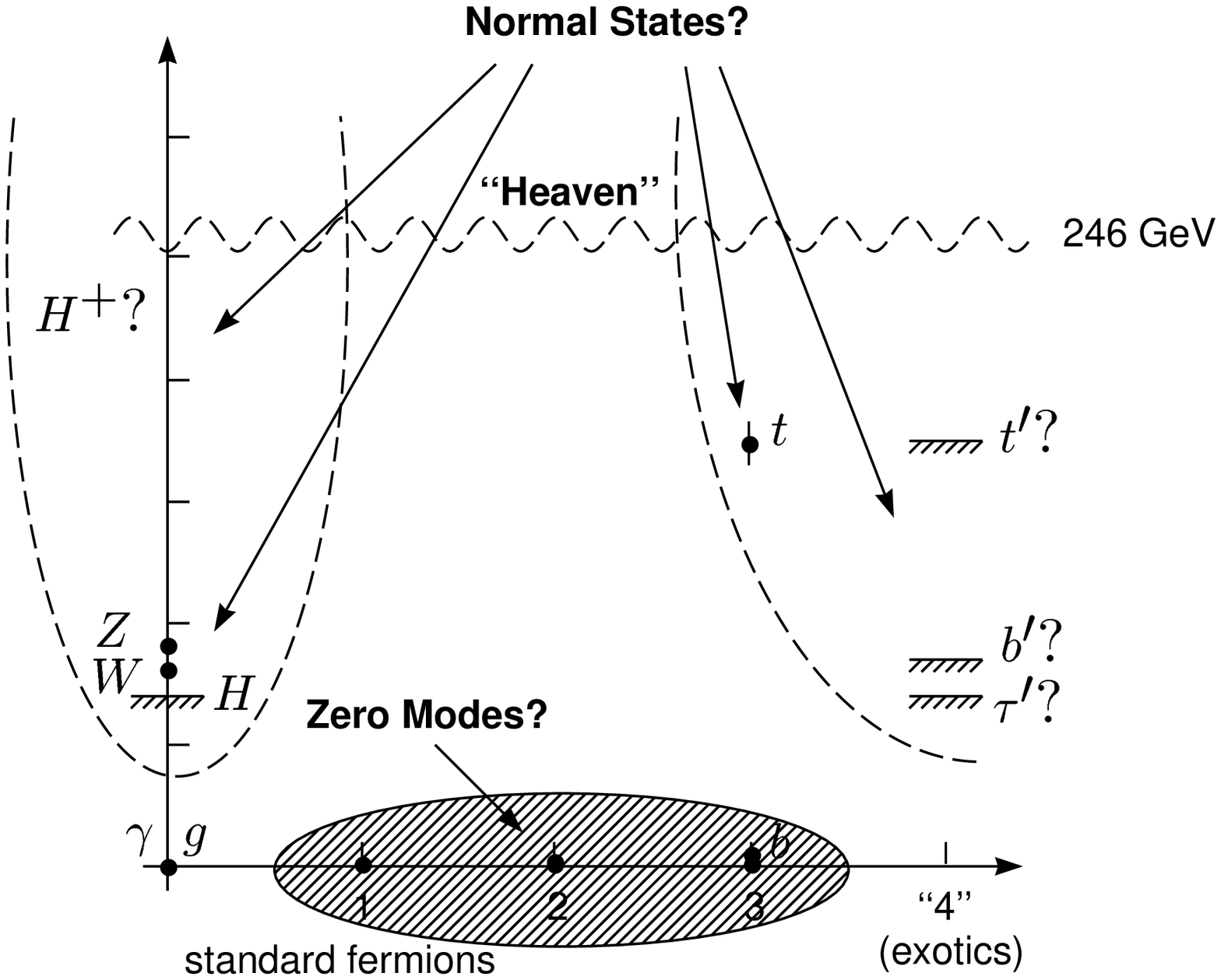}}}\fi
%
\vskip -0.2cm
\caption{Perspective on scale of mass generation.}
\end{figure*}

\vskip 0.5cm
\noindent{\large\bf II. ``Fritzsch'' Path: from {$\sin\theta_C \simeq \sqrt{m_d/m_s}$}}
\vskip 0.3cm

The Gatto--Sartori--Tonin/Cabibbo--Maiani \cite{GSTCM} 
empirical relation given above was
 ``explained" by the Ansatz (or, texture) of Weinberg \cite{Wein},
and generalized by Fritzsch \cite{Fritzsch}  to $3\times 3$ form 
soon after the discovery of the $\Upsilon$, \vskip -0.4cm
\begin{equation}
M_d = \left[ \begin{array}{cc}
          0  &  a \\
          a  &  b
          \end{array} \right]
\ \ \ \ \Longrightarrow \ \ \ \ 
M_{u,d} = \left[ \begin{array}{ccc}
          0 & a & 0 \\
          a & 0 & b \\
          0 & b & c
          \end{array} \right].
\label{eq:FritzM}
\end{equation} 
Assuming $M_u$, $M_d$ to be hermitian for simplicity,
they are diagonalized by unitary transforms
$U^\dagger M_u U$, $D^\dagger M_d D$.
In terms of small mass ratios, one has \vskip -0.4cm
\begin{equation}
D \simeq \left[ \begin{array}{ccc}
          1 & \sqrt{d\over s} & \sqrt{d\over b} {s\over b} \\
          - \sqrt{d\over s} & 1 & - \sqrt{s\over b} \\
          - \sqrt{d\over b} &  \sqrt{s\over b} & 1
          \end{array} \right],
\label{eq:D}
\end{equation}
and likewise for $U$.
Putting back the phases, the Kobayashi-Maskawa matrix \cite{KM} is
the difference between $U$ and $D$, that is
$V \cong U^T(1,\  e^{i\sigma},\ e^{i\tau}) D$, hence \vskip -0.6cm
\begin{equation}
V \simeq \left[ \begin{array}{ccc}
    1 & \sqrt{d\over s}- \sqrt{u\over c}\, e^{i\sigma} & 
     \sqrt{s\over b}\sqrt{u\over c}\, e^{i\sigma} - \sqrt{u\over t}\, e^{i\tau}
                                                                             + \sqrt{d\over b} {s\over b} \\
    -\sqrt{d\over s} + \sqrt{u\over c}\, e^{-i\sigma} & 
    1 & - \sqrt{s\over b} + \sqrt{c\over t}\, e^{i(\tau - \sigma)}\\
   \sqrt{d\over b} + \sqrt{d\over s}\sqrt{c\over t}\, e^{i(\sigma - \tau)} &  
     \sqrt{s\over b} - \sqrt{c\over t}\, e^{i(\sigma - \tau)} & 1
          \end{array} \right].
\label{eq:FritzV}
\end{equation}
One has the approximate relations
$V_{ub}/V_{cb} \sim \sqrt{u/c}$,
$V_{td}/V_{ts} \sim \sqrt{d/s}$, hence the smallness of
$V_{ub}$ compared to $V_{cb}$ comes out quite naturally.

In 1978, bottom had just been discovered (in the form of $\Upsilon$),
but the top quark mass was not known at all.
Fritzsch had hoped to predict $m_t$ from $V_{cb}$,
or {\it vice versa}, by tuning the cancellations for $V_{cb}$,
which is highlighted in   (\ref{eq:FritzV}).
The chain of colliders, PEP $\to$ PETRA $\to$ TRISTAN $\to$ LEP,
were largely aimed at discovering the top quark, 
a hope vanquished by the ARGUS discovery of large $B$-$\bar B$ mixing.
Today, the top quark mass has been measured at the Tevatron to be very,
very heavy, completing the mass ratios
\begin{eqnarray}
{d\over s} \simeq {1\over 21} - {1\over 18} &\gg& 
{u\over c} \sim {1\over 390} - {1\over 200}, \\
{s\over b} \simeq {1\over 40} - {1\over 25}  &\gg& 
{c\over t} \sim {1\over 280}.
\end{eqnarray}
With such heavy top,
does the Fritzsch Ansatz of   (\ref{eq:FritzV}) still work?

\vskip 0.3cm
\noindent{\bf  \underline{Critique}: \rm Main Problem from $V_{cb} \ll V_{us}$ }
\vskip 0.2cm

With $V_{us} \simeq \sqrt{d/s}$
and $V_{cb} \simeq \sqrt{s/b} - \sqrt{c/t}\, e^{i\phi}$,
since $\sqrt{s/b} \sim\sqrt{d/s}$, there are serious flaws 
even without $m_t$ being extremely heavy.
Below is a critique \cite{HeHou}:
\begin{enumerate}
\item[(a)] Need $\vert\tau - \sigma \vert 
{\ \lower-1.2pt\vbox{\hbox{\rlap{$<$}\lower5pt\vbox{\hbox{$\sim$}}}}\ } 
10^\circ - 15^\circ$ to allow for cancellation.

\item[(b)] $m_t$ {\it tuned} to cancel  $\Longrightarrow 
m_t {\ \lower-1.2pt\vbox{\hbox{\rlap{$<$}\lower5pt\vbox{\hbox{$\sim$}}}}\ }
60$ GeV. Is this arbitrary, or predictive?

\item[(c)] \underline{$\sigma \approx\pm \pi/2$}  $\longmapsto
\vert V_{us} \vert^2 \sim d/s + u/c$\\
\phantom{xxx} But combine with (a) $\Longrightarrow$ CP violating phase {\it not} predicted.
 
\item[(d)] And, because of (b): RULED OUT WHEN $m_t > M_W$.\\
\phantom{xxx} Impossible to account for 
$V_{cb} \ll \sqrt{s/b} \sim V_{us} \, \nearrow$
\end{enumerate}

Thus, the Fritzsch Ansatz was not very satisfactory in the first place.

\vskip 0.3cm
\noindent{\bf \underline{He-Hou Variant}: \rm Relating $\tau_b \propto 1/m_t$ }
\vskip 0.2cm

Dimopoulos once gave a talk at CERN and dubbed
``He(e)-H(o)u did it."
Well,
it really should be He[r]--Ho[u] phonetically.
There is an interesting story about how it got started:
He[r]--Ho[u] did it in Spring 1989 over some L\" owenbra\" u (beer)
sitting on Ludwigstrasse in M\" unchen,
not far from the Ludwig--Maxmilians--Universit\" at where Fritzsch is.
Say we were inspired \ldots

We started by discussing the culprit:
$\sqrt{s/b}$ was way too large for $V_{cb}$ in (\ref{eq:FritzV}).
We said to ourselves: ``{\it Just Dump it!}",
and went on with more beer \ldots

More sober the morning after, we made the 
\vskip 0.2cm
\centerline{\fbox{\bf Postulate: No $s$-$b$ or $d$-$b$ mixing}
\ $\Longrightarrow$ \ one rotates only in $d$-$s$ plane.}
\vskip 0.2cm
\noindent With $M_u$ unchanged from   (\ref{eq:FritzM}), 
our Ansatz amounted to \vskip-0.4cm
\begin{equation}
M_{d} = \left[ \begin{array}{ccc}
          0 & a & 0 \\
          a & b & 0 \\
          0 & 0 & c
          \end{array} \right].
\label{eq:HeHouM}
\end{equation}
One immediately finds that 
(i) the phase $\tau$ could now be absorb into redefining
 $b_L$ and $b_R$ and
(ii) one just sets $s/b$ and $d/b \rightarrow 0$ 
in   (\ref{eq:FritzV}) and arrive at  \vskip-0.4cm
\begin{equation}
V \simeq \left[ \begin{array}{ccc}
    1 & \sqrt{d\over s}- \sqrt{u\over c}\, e^{i\sigma} & - \sqrt{u\over t}\, e^{i\sigma} \\
    -\sqrt{d\over s} + \sqrt{u\over c}\, e^{-i\sigma} &1 & \sqrt{c\over t}  \\
     \sqrt{d\over s}\sqrt{c\over t} & - \sqrt{c\over t} & 1
          \end{array} \right],
\label{eq:HeHouV}
\end{equation}
which is considerably simpler.
The salient features are:
\begin{enumerate}
\item[(a)] \underline{$V_{cb} = \sqrt{c/t}$}: We related
Long $\tau_b$ (Smallness of $V_{cb}$) $\Longleftrightarrow$
Heaviness of $m_t$
without double fine-tuning as in the Fritzsch case.

Using QCD running only and
ignoring the potentially more important weak corrections 
(running of $\lambda_t$), we found that 
\begin{equation}
0.03{\ \lower-1.2pt\vbox{\hbox{\rlap{$<$}\lower5pt\vbox{\hbox{$\sim$}}}}\ }
\vert V_{cb}\vert
{\ \lower-1.2pt\vbox{\hbox{\rlap{$<$}\lower5pt\vbox{\hbox{$\sim$}}}}\ }0.06
\ \ \ \Longrightarrow \ \ 750 \mbox{GeV} 
{\ \lower-1.2pt\vbox{\hbox{\rlap{$>$}\lower5pt\vbox{\hbox{$\sim$}}}}\ }
m_t {\ \lower-1.2pt\vbox{\hbox{\rlap{$>$}\lower5pt\vbox{\hbox{$\sim$}}}}\ } 
200 \mbox{GeV}.
\label{eq:HeHoumt}
\end{equation}
This was a tantalizing result, but also indicative of the problem
that we should not have ignored the running of $\lambda_t$.

\item[(b)] $\vert V_{ub}/V_{cb}\vert = \sqrt{u/c}$ 
and $\vert V_{td}/V_{ts} \vert = \sqrt{d/s}$ become exact. 
Only $V_{cb}$ is related to $1/m_t$,
the rest of $V_{\rm KM}$ is expressed in terms of $V_{cb}$ and 
ratios of masses of first two generations.
The prediction of very small $V_{ub}$
comes out naturally!

\item[(c)] \underline{$\sigma \approx\pm \pi/2$} as in Fritzsch.

{\it But}, there is {\it only one single phase} $\longrightarrow$
 CP violating phase determined.
 
\item[(d)] \underline{Maximal CP}: Strictly! That is, from \vskip -1.1cm
\begin{eqnarray}
J_{\rm CP} &\propto& \mbox{Im}(V_{us} V_{ub}^* V_{cs}^* V_{cb})
\ \simeq \ \sqrt{m_d\over m_s} \sqrt{m_u\over m_t} 
\sqrt{m_c\over m_t} \left(1 + {m_u\over m_c}\right)\, \sin\sigma
  \nonumber \\  
&\simeq & \vert V_{us} \vert\, \vert V_{ub}\vert 
\, \vert  V_{cs}\vert\, \vert  V_{cb}\vert\, \sin\sigma 
\ \cong \ (3\mbox{--}4)  \times 10^{-5},
\end{eqnarray}
the CP phase is clearly maximal if $\sigma = \pm \pi/2$.

\item[(e)] Wolfenstein expansion \cite{Wolf}: 
Take $\sin\theta \equiv \lambda$, $V_{cb} \sim \lambda^2$,
and $V_{ub}/V_{cb} = \sqrt{u/c} \sim \lambda^2$, after some phase redefinitions,
we find the Wolfenstein form \vskip -0.5cm
\begin{equation}
V= \left[ \begin{array}{ccc}
          \ 1 \ & \lambda \ & \ \lambda^5 \mp i\lambda^4 \\
          \ -\lambda\ & \ 1 \ & \ \lambda^2 \\
          \ (1 \mp i\lambda) \lambda^3 \ & \ -\lambda^2 \ & \ 1
          \end{array} \right].
\label{eq:HeHouWolf}
\end{equation}
\end{enumerate}
\vskip -0.3cm
The upshot is that we have a completely predictive Ansatz:
{\it heavy top, maximal CP, {\rm and} 
$V_{\rm KM}$ fixed in terms of $\sqrt{m_i/m_j}$ ratios!}
The mass matrices of $M_d$ in   (\ref{eq:HeHouM})
and $M_u$ in   (\ref{eq:FritzM}) were just (unkown to us!)
the Georgi and Jarlskog (``texture") form \cite{GJ},
hence the work anticipated the Top $\searrow$ Down
approach of SO$(10)$ SUSY-GUT
models of Dimopoulos, Hall and Raby \cite{DHR},
which can be viewed as a high scale approach 
to remedy the over-heavy top.

As $V_{cb}$ went down from $\sim 0.06$ in 1989 to $\sim 0.04$ by 1993,
it became very difficult even for (\ref{eq:HeHouV})
to account for the smallness of $V_{cb}$.
The problem remains severe even for 
the SO$(10)$ based SUSY-GUT models.

\vskip 0.8cm
\noindent{\large\bf III. ``Wolfenstein'' Path: 
from $\vert V_{cb}\vert \simeq \vert V_{us}\vert^2 \ll 1$}
\vskip 0.3cm

As we have mentioned the Wolfenstein parametrization in 
  (\ref{eq:HeHouWolf}) already, it is useful to gain some 
historic perspective.

\vskip 0.3cm
\noindent{\bf \underline{Then and Now}: \rm $B$ Decay Data from 1983 $\leadsto$ 1994}
\vskip 0.2cm

It was discovered in 1983 that the 
$B$ lifetime was much longer than expected.
Together with the absence of $b\to u$ transitions, 
the new experimental values were
\begin{equation}
V_{cb} \approx 0.06,\ \ \ \ \vert V_{ub}/V_{cb} \vert < 0.2,
\label{eq:old}
\end{equation}
indicating that $V_{ub}^2 \ll V_{cb}^2 \ll V_{us}^2$.
Taking note of this,
Wolfenstein suggested that \cite{Wolf}
$\lambda \equiv V_{us} \simeq 0.22$ could be
taken as an expansion parameter for $V_{\rm KM}$,
and proposed to parametrize the KM matrix as  \vskip -0.4cm
\begin{equation}
V= \left[ \begin{array}{ccc}
          \ 1-{1\over 2}\lambda^2 \ & \lambda \ & \
          A\lambda^3(\rho-i\eta) \\
          \ -\lambda\ & \
          \ 1-{1\over 2}\lambda^2 \ & \
          \ A\lambda^2 \\
          \ A\lambda^3(1-\rho-i\eta) \ & \ -A\lambda^2 \ & \ 1
          \end{array} \right],
\label{eq:Wolf}
\end{equation}
to order $\lambda^3$, 
with $A \approx 5/4$ and $\rho^2 + \eta^2 <1$ from (\ref{eq:old}).
The parametrization has since become a reference standard \cite{PDG}
because of its value as a mnemonic device, especially in regards the
popular unitarity triangle representation for CP violation,

Fourteen years has elpased since Wolfenstein's proposal of 
  (\ref{eq:Wolf}).
With the advent of CLEO II data 
and the development of HQET,
the values for $V_{cb}$ and $\vert V_{ub}/V_{cb} \vert$
have been consistently dropping in the past few years \cite{PDG},
stabilizing more or less by 1994. 
The current values are
\begin{equation}
\vert V_{cb} \vert = 0.040 \pm 0.005,
                                                                         \ \ \ \
\vert V_{ub}/V_{cb} \vert = 0.08 \pm 0.02.
\label{eq:new}
\end{equation}
Thus, we now have
\begin{equation}
A = 0.8 \pm 0.1,
\label{eq:A}
\end{equation}
which is
{\it down by 1/3} compared to ten years ago.
But the drop in $\vert V_{ub}\vert$ is more dramatic:
{\bf a factor of 4 down} from that of   (\ref{eq:old}), giving
\begin{equation}
\vert V_{ub}\vert = 0.0032 \pm 0.0009.
\label{eq:Vub}
\end{equation}

A factor of 1/4 corresponds to $\sim \lambda$ suppression.
Noting that $\lambda \cong 0.2205$
hence $\lambda^2 \cong 0.0486$, $\lambda^3 \cong 0.0107$ 
and $\lambda^4 \simeq 0.0024$,
we can now define \cite{HW}
\begin{equation}
V_{ub} \equiv A\lambda^4(\rho^\prime - i \eta^\prime) \equiv B\lambda^4 e^{-i\phi},
\label{eq:rho'}
\end{equation}
where $\rho \equiv \rho^\prime\lambda$, $\eta \equiv \eta^\prime\lambda$
(i.e. $d\rho/d\lambda = \rho^\prime$ as a memory device)
and $\phi \equiv \tan^{-1} \eta^\prime/\rho^\prime$.
Numerically,
\begin{equation}
B \equiv A\sqrt{\rho^{\prime 2} + \eta^{\prime 2}} = 1.3 \pm 0.5,
\label{eq:B}
\end{equation}
and interestingly, $A B \sim 1$.
We note that the original Wolfenstein parametrization is still better suited
as a mnemonic device (see Fig. 2), 
but the change in powers of $\lambda$ for $V_{ub}$ 
might have some significance.

\begin{figure*}[htb]
\let\picnaturalsize=N
\def\picsize{2.0in}
\def\picfilename{shizuoka2.eps}
\ifx\nopictures Y\else{\ifx\epsfloaded Y\else\input epsf \fi
\global\let\epsfloaded=Y
%
%
\centerline{\ifx\picnaturalsize N\epsfxsize \picsize\fi \epsfbox{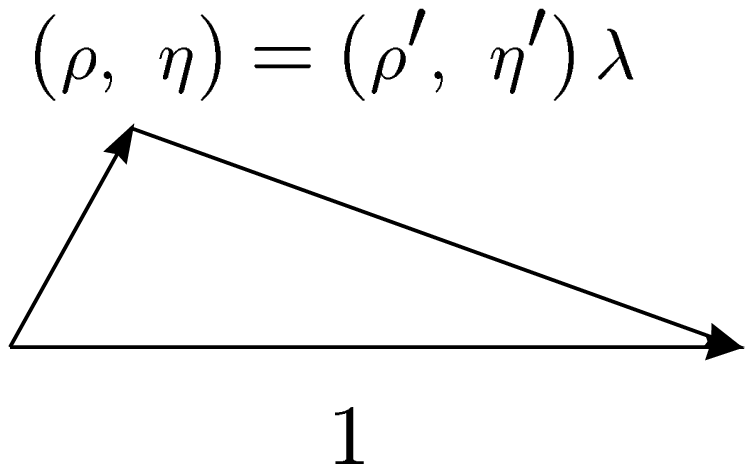}}}\fi
%
\vskip -0.2cm
\caption{Unitarity triangle illustrating the change
$V_{ub}: \lambda^3 \leadsto \lambda^4$.}
\end{figure*}

Whether this is mere numerology or not
depends on how you take the original Wolfenstein paper.
The original proposal emphasized 
a possible $\lambda$-expansion \cite{Wolf},
a point which was largely lost because of 
its great  success as a mnemonic device.
Afterall, Wolfenstein freely admitted that $\lambda \simeq 0.22$
is not so small as an expansion parameter, making the order of the series,
if any, ambiguous.

If one takes the series expansion idea seriously \cite{HW},
however, the change in $V_{ub}: \lambda^3 \leadsto \lambda^4$
amounts to a change in order of $\lambda$
from $odd \leadsto even$.
This may have profound implications for possible underlying dynamics
that relates masses and mixing angles.
With the recent measurement of $m_t(m_t)  \cong 175$ GeV,
it is intriguing to note that, 
as shown in Eqs. (4) and (5) we now have
\begin{equation} \label{eq:mratio}
          {m_d\over m_s}  \sim  {m_s\over m_b} \sim \lambda^2
                                               \gg         
          {m_u\over m_c}  \sim  {m_c\over m_t} \sim \lambda^4.
\end{equation}
The relations suggest that 
$V_{\rm KM}$ comes mostly from down quark sector,
while the up-type quark sector
contributes only at higher order in $\lambda$.
Let us therefore explore along these lines in more detail.

\vskip 0.3cm
\noindent{\bf \underline{``Gestalt" Switch}: 
\rm Hint from \fbox{$V_{ub}: \lambda^3 \longmapsto \lambda^4$}}
\vskip 0.2cm

Combining Fritzsch with Wolfenstein,
we have the old empirical relation
\begin{equation}
V_{us} = \lambda \simeq \sqrt{m_d\over m_s}.
\label{eq:Wein}
\end{equation}
With the falling $V_{cb}$, several authors have noted \cite{sob,Ng} that
the relation 
\begin{equation}
V_{cb} \equiv A\lambda^2 \simeq {m_s\over m_b},
\label{eq:sob}
\end{equation}
now holds rather well.
It appears then that 
{\it the KM matrix $V$ is
mostly due to the down type sector, while
eqs. (\ref{eq:rho'}) and (\ref{eq:mratio}) suggest that perhaps $V_{ub}$ 
 --- hence $CP$ violation --- originates from the up type sector}.
We thus reach the Proposition:
\begin{equation}
D\sim V_{\rm KM}.
\label{eq:Prop}
\end{equation}
Let us see how far we can go with this thought.

For simplicity, let us assume hermitian mass matrices
(this limits the form of the possible underlying dynamics).
Ignoring the $u$-type quark mass matrix to first approximation 
since its mass ratios are subdominant,
we can reconstruct the down type quark mass matrix 
by rotating back with the KM matrix, that is \vskip -0.4cm
\begin{equation}
M_d \cong V \left[ \begin{array}{ccc}
                         -m_d  &   0      &  0 \\
                         0        &  m_s  &  0 \\
                         0        &   0      &  m_b \\
          \end{array} \right] V^\dagger.
\label{eq:Ma}
\end{equation}
\begin{itemize}
\item Approximate ``Texture Zero":  \fbox{$(M_d)_{11} = 0$}

The empirical relation of   (\ref{eq:Wein}) can be maintained
IF \cite{Ma} $(M_d)_{11} = 0$ in   (\ref{eq:Ma}), 
which is nothing but the old texture zero of Weinberg.
It holds approximately true numerically, namely \vskip -0.5cm
\begin{equation}
(M_d)_{11} = - m_d \vert V_{ud}\vert^2
                          + m_s \vert V_{us}\vert^2 + m_b \vert V_{ub}\vert^2 \cong 0,
\end{equation}
as can be most easily checked by making a series expansion in $\lambda$.

[In 1991, one still had $V_{ub} \sim \lambda^3$.
Ma attempted, therefore, at a second zero,
$(M_dM_d^\dagger)_{12} \simeq 0$,
suggesting a second relation \cite{Ma}
$m_s^2/m_b^2 = -V_{ub} V_{cb}/V_{us}$.
Both were of order $\lambda^4$ in 1991.
With   (\ref{eq:new}) hence   (\ref{eq:rho'}), however, 
the righthand side is now of order $\lambda^5$,
and the relation no longer holds.]
\item Idealize: {\it Least Number of Parameters}

To maintain the relation of    (\ref{eq:sob}),
without loss of generality, let us
redefine quark fields to make $M_d$ real symmetric.
This relegates all CP violating effects to the up-type sector,
{\it i.e.} $V_{ub}$ is from $M_u$.
We therefore idealize our proposition of   (\ref{eq:Prop}) and \vskip -0.5cm
\begin{equation}
\mbox{ {\bf Postulate: }} \fbox{$D_L \equiv V_{\rm KM}(V_{ub} = 0)$},
\label{eq:DL}
\end{equation}
which corresponds to the {\it least number of parameters
needed to account for both $d$-type quark masses and
$V$, with $V_{ub} = 0$}.
There are just two parameters, $\lambda$ and $A$, 
or, we define $\lambda \equiv V_{us}$,
$\delta \equiv V_{cb} \cong A\lambda^2$, \vskip -0.5cm
\begin{equation}
D_L = \left[ \begin{array}{ccc}
           \sqrt{1-\lambda^2}    &    \lambda    &     \fbox{0} \\
           -\lambda \sqrt{1-\delta^2}   &
           \sqrt{1-\lambda^2} \sqrt{1-\delta^2} &  \delta \\
           \lambda\delta    &     - \sqrt{1-\lambda^2}\, \delta 
                                      &     \sqrt{1-\delta^2} 
          \end{array} \right].
\label{eq:DL2}
\end{equation}
With $(D_L)_{13} = 0$ by fiat, Ma's observation 
is now reformulated as
\begin{eqnarray}
(M_d)_{11} &=& - m_d (D_L)_{11}^2
                          + m_s (D_L)_{12}^2  \nonumber \\
           &=& - m_d\, (1-\lambda^2) + m_s\, \lambda^2 = 0.
\label{eq:Mad}
\end{eqnarray}
Enforcing this ``texture" zero strictly gives \vskip -0.6cm
\begin{equation}
{m_d\over m_s} = {\lambda^2\over 1-\lambda^2},
\label{eq:dos}
\end{equation}
to all orders in $\lambda$, leading to the relation \vskip -0.6cm
\begin{equation}
 V_{us} \equiv \lambda = \sqrt{m_d\over m_s + m_d}  \cong s_{12},
\label{eq:lambda}
\end{equation}
or $\tan \theta_{12} = \sqrt{m_d/m_s}$.
We see that the idealization (\ref{eq:DL2}) of  our proposition (\ref{eq:Prop}) 
leads to a refinement of  Eq. (\ref{eq:Wein}).
\item Second ``Texture Zero":   \fbox{$(M_d)_{13} = 0$}

Coming back to relation (\ref{eq:sob}), we note that it implies
$m_s\ : \ m_b = A\lambda^2\ : \ 1$, while the all orders relation 
(\ref{eq:dos}) gives $m_d\ : \ m_s = \lambda^2\ : \ 1 - \lambda^2$.
Taken together they suggest that 
\begin{equation}
-m_d\ : \ m_s \ : \ m_b = -A\lambda^4\ : \ A\lambda^2 - A\lambda^4\ : \ 1
,
\label{eq:Mdratio}
\end{equation}
Of course, we are not sure of corrections to the relation (\ref{eq:sob})
since $m_s$, as well as $m_b$, are less precisely known
as $V_{cb}$. But if we take the ratio (\ref{eq:Mdratio}) as is
and idealize, we get an additional texture zero ($M_d$ is real symmetric).
Define $\hat M_d = M_d/m_b$, multiplying out    (\ref{eq:Ma}), 
we find  \vskip -0.6cm
\begin{equation}
\hat M_d = \left[ \begin{array}{ccc}
              0 & A\lambda^3 & 0 \\
              A\lambda^3 & A\lambda^2+(A-2)A\lambda^4 &
              A\lambda^2-A^2 \lambda^4 \\
              0 & A\lambda^2-A^2\lambda^4 &  1 - A^2\lambda^4
             \end{array} \right].
\label{eq:Md}
\end{equation}
The texture zero of $(M_d)_{13} = 0$ implies the relation \vskip -0.3cm
\begin{equation}
V_{cb} \equiv A\lambda^2 = {m_s + m_d\over m_b} \cong s_{23},
\label{eq:Al2}
\end{equation}
which modifies Eq. (\ref{eq:sob}) slightly, 
but is less convincing than (\ref{eq:lambda}).
\end{itemize}

Thus, with the three parameters $m_b$ (scale parameter), 
$A$ ($\sim 1$) and $\lambda$, one can account for 
the 5 parameters $m_d$, $m_s$, $m_b$ and 
$\vert V_{us}\vert$, $\vert V_{cb}\vert$, resulting in two relations,
Eqs. (\ref{eq:lambda}) and (\ref{eq:Al2}).

\vskip 0.3cm
\noindent{\bf \underline{Up--type Sector}: 
\rm CP Violation, $m_u$, and $V_{ub}$}
\vskip 0.2cm

Let us extend the previous arguments to the up-type sector.
With the postulate of  (\ref{eq:DL}) that $D_L$ accounts for
$V_{\rm KM}$ up to $V_{ub} = 0$, we simply have \vskip -0.4cm
\begin{equation}
U_L \equiv D_L V^\dagger
     \simeq \left[ \begin{array}{ccc}
          1   &   0   &  -B \lambda^4 e^{-i\phi}  \\
           0   &  1   &   0 \\
           B \lambda^4 e^{i\phi} & 0 & 1 
          \end{array} \right] + {\cal O}(\lambda^6).
\label{eq:UL}
\end{equation}
Taking $M_u$ to be hermitian also, we have,
analogous to Eq. (\ref{eq:Ma}),
\begin{equation}
\hat M_u = U_L  \left[ \begin{array}{ccc}
                         -\hat m_u  &   0      &  0 \\
                         0        &  \hat m_c  &  0 \\
                         0        &   0      &  \hat m_t 
          \end{array} \right]  U_L^\dagger,
\label{eq:Mau}
\end{equation}
where we normalize with respect to $m_t$,
{\it i.e.} $\hat m_t = 1$.
Noting the approximate orders
$m_t : m_c : m_u \approx 1 : \lambda^4 : \lambda^8$
from (\ref{eq:mratio}),
we multiply out  (\ref{eq:Mau}) and get
\begin{equation}
\hat M_u =  \left[ \begin{array}{ccc}
                         -\hat m_u +B^2\lambda^8 &  0 
                            &  -B \lambda^4 e^{-i\phi} \\
                         0 &  \hat m_c  &  0 \\
                         -B \lambda^4 e^{i\phi} &  0 
                           &  1 
          \end{array} \right] + {\cal O}(\lambda^{10}).
\end{equation}
We can now proceed to idealize.

\begin{itemize}
\item ``Texture Zero":  $(M_u)_{11} = 0$ analogy

The $(\hat M_u)_{11}$ term is at order $\lambda^8$
and can be removed by envoking a texture zero condition
analogous to but weaker than (\ref{eq:Mad}),
\begin{equation}
(M_u)_{11} = - m_u \vert U_L\vert_{11}^2
                          + m_t \vert U_L\vert_{13}^2 = 0,
\end{equation}
which holds at least up to order $\lambda^{12}$.
This leads to the relation
\begin{equation}
\vert V_{ub}\vert \equiv B\lambda^4 \cong \sqrt{m_u\over m_t},
\label{eq:Bl4}
\end{equation}
hence {\it both $V_{ub}$ and $m_u$ are generated via
diagonalizing the $u$-$t$ mixing element},
analogous to the original
$\sin\theta_C \simeq \sqrt{d/s}$ relation.

\item Reduction of Parameters: {\it Least Number}

Eq. (\ref{eq:mratio}) suggest that we could write  $\hat m_c = C\lambda^4$ with
$C \simeq 1$, hence 
\begin{equation}
\hat M_u \cong
          \left[ \begin{array}{ccc}
           0 & 0 & \ -B \lambda^4 e^{-i\phi} \\
            0 & C\lambda^4 & 0 \\
          -B \lambda^4 e^{i\phi} & 0 & 1
          \end{array} \right],
\label{eq:Mu}
\end{equation}
where  the $11$ element is  at ${\cal O}(\lambda^{12})$,
while the other $0$'s are at ${\cal O}(\lambda^{10})$.
Since both $C$ and $B$ are $\sim 1$,  setting  $C = B$
reduces the number of parameters, 
leading to a second, ``geometric" relation
\begin{equation}
{m_u\over m_c}={m_c\over m_t}\ =B\lambda^4.
\label{eq:geo}
\end{equation}
We are now really deep into numerology.
Since from (\ref{eq:A})  and (\ref{eq:B}) we have
$A \simeq 1 - \lambda$ and $B \simeq 1 + \lambda$,
it may well be that $B =1/A$ or $1$ in Nature,
such that 
$m_u/m_c = m_c/m_t = \vert V_{ub}\vert = \lambda^4$
or $A^{-1}\lambda^4$,
and just {\it two} additional parameters, 
$m_t$ and $\phi$,
{\it might} account for the remaining {\it five} parameters
$m_u$, $m_c$, $m_t$ and $V_{ub} = \vert V_{ub}\vert e^{-i\phi}$.
Given the uncertainties in mixing, 
and especially in the {\it lighter} quark masses \cite{PDG},
these reults are not inconsistent with data!
Note that the CP violating phase $\phi = \tan^{-1}(\eta^\prime/\rho^\prime)$
is a free parameter, and does not arise from mass ratios.

\item Naturalness of $V_{ub}$ ({\it and CP violation}) from $M_u$

Eqs. (\ref{eq:rho'}--\ref{eq:sob}) 
imply that 
$m_d/m_b = A\lambda^4 \ltap m_u/m_c \sim m_c/m_t
 \sim \vert V_{ub}\vert \sim B\lambda^4$, 
so we could arrange to have $V_{ub}$ to arise from $\hat M_d$
while $\hat M_u \simeq \mbox{diag}\,(-\lambda^8,\ \lambda^4,\ 1)$
to be already diagonal.
This corresponds to taking the off-diagonal piece from $U_L$
in   (\ref{eq:UL}) and placing it in $D_L$.
However, this is {\it unnatural} since if
$V_{ub}$ comes from $M_d$ and starts at ${\cal O}(m_d/m_b)$,
then \vskip -0.5cm
\[
\left\{ \begin{array}{l}
           V_{ub} \propto m_d/m_b \\
          m_t\ \mbox{arbitray and large}
          \end{array}
\right. \ \ \ \ \ \ \ \Longrightarrow \varepsilon_K\ \mbox{easily fails!}
\]
It is therefore important, as in (\ref{eq:UL}) or (\ref{eq:Mu}), to couple 

\ \ \ \ \ {\it Smallness of $CP$ Violation Effects $\Longleftrightarrow$
Heaviness of the Top Quark},

like in (\ref{eq:geo}), where $\vert V_{ub} \vert \sim \sqrt{u/t} \sim c/t$,
leading to \vskip -0.5cm
\begin{equation}\label{eK}
\varepsilon_K \propto \mbox{Im}(V_{ub})^2\, m_t^2,
\end{equation}
which would be stable against the rise of $m_t$.
The placement of CP phase in $(M_u)_{13}$ may 
therefore be rooted in dynamics.
\end{itemize}

\vskip 0.3cm
\noindent{\bf \underline{Discussion}
}
\vskip 0.2cm

\begin{itemize}
\item Since we are just applying the $\lambda$-expansion,  
the phenomenology is the same as in SM. 
In particular, 
the invariant area of 
the unitarity triangle
\begin{equation}
J_{CP} = \mbox{\rm Im}(V_{us} V_{ub}^\ast V_{cs}^\ast V_{cb})
\simeq A^2 \eta^\prime \lambda^7 \ltap 3\times 10^{-5},
\end{equation}
is of order $\lambda^7$.
Though it is still convenient to use $\rho$ and
$\eta$ of Wolfenstein \cite{PDG} for the unitarity triangle,
but with $\rho$, $\eta = \rho^\prime\lambda$, $\eta^\prime\lambda$,
and $\rho^\prime,\ \eta^\prime \sim 1$
the unitarity triangle appears a bit squashed (see Fig. 2).
The CP phase angle $\phi \equiv \tan^{-1} \eta^\prime/\rho^\prime$
remains a free parameter.
\item $\delta$-expansion {\it vs.} $\lambda$-expansion

It would be appealing if some symmetry
or dynamical mechanism underlies the 
possible reduction of 2 to 5 parameters
from the 10 quark masses and mixing angles. 
Discrete symmetries \` a la Ma \cite{Ma}
can be constructed by adding extra Higgs doublets.
However,
these usually do not add insight to the $\lambda^n$ power behavior
for mixing angles and mass ratios.
Note that eqs. (\ref{eq:Md}) and (\ref{eq:Mu}) suggest
an expansion in even powers of $\lambda$,
{\it except} for $(\hat M_d)_{12} = (\hat M_d)_{21} \cong \lambda^3$.
This is because,
with $V_{ub}$ changed from order $\lambda^3$
to order $\lambda^4$,
the only term odd in $\lambda$ is just 
$\vert V_{us}\vert = \vert V_{cd}\vert$ itself.
Defining as before $\delta \equiv V_{cb} \equiv A\lambda^2$, 
we find, to leading order in $\delta$, \vskip -0.6cm
\begin{equation}
\hat M_d \cong \left[ \begin{array}{ccc}
              0 & \delta^{3/2} &  0 \\
              \delta^{3/2} & \delta &  \delta \\
              0 & \delta & 1
             \end{array} \right], \ \ \ \
\hat M_u \simeq
          \left[ \begin{array}{ccc}
          0 & 0 & -\delta^2\, e^{-i\phi} \\
          0 & \delta^2 & 0 \\
          -\delta^2\, e^{i\phi} & 0 & 1
          \end{array} \right].
\label{eq:ansatz}
\end{equation}
So perhaps $\delta \simeq 1/20 - 1/30$ is 
the actual expansion parameter.
The Wolfenstein's $\lambda = \sqrt{\delta/A}$
is empirically tied to $(\hat M_d)_{11} = 0$ \cite{Ma},
but one needs to understand the relatively large ratio $m_d/m_s$.
\item The 1991 ansatz of Ng and Ng \cite{Ng}
also starts with the approximate relations of 
(\ref{eq:mratio}), but they impose $(M_d)_{32} = 0$,
which we find unnecessary.
The 1992 Ansatz of Giudice \cite{Giudice}
starts fromSUSY GUTS (Georgi--Jarlskog texture),
and relates $M_d$ and $M_e$ with {\it ad hoc} assumptions,
and arbitrarily sets $(M_d)_{23} = 2\,(M_d)_{22}$.
Both of these works do not utilize the $\lambda$-expansion.

\item General Texture Analysis

In 1993 Ramond, Roberts and Ross \cite{RRR}
made a systematic extension of the Dimopoulos--Hall--Raby approach.
They studied all the possible {\it texture zeros}
(which implies mass-mixing relations)
of symmetric or hermitian $M_u$ and $M_d$ matrices
{\it at SUSY GUT scale},
then evolved them down from $\Lambda_{\rm GUT}$ to
$\Lambda_{\rm Weak}$.
They find 5 textures, the third of which is close to ours
(see their Tables 1 and 2).
They do make a $\lambda$-expansion in a special way,
with $\sqrt{\rho^2 + \eta^2} \sim \lambda$.
They view $\lambda$ as possibly emerging from ratios of
v.e.v.'s.
But because the textures exist at GUT scale,
they have the same old generic problem of DHR,
namely it is difficult to bring $V_{cb}$ below 0.045 (see their Table 3).

\item Weak Scale Mass Generation?

In comparison, our Ansatz works BELOW WEAK SCALE,
without any attempt to evolve upwards.
We find   (\ref{eq:ansatz}) to be suggestive of an underlying 
radiative mechanism, perhaps not far above the electroweak scale \cite{radm}.
For example, the parameter $\delta\equiv A\lambda^2$
defined in eq. (\ref{eq:DL}) could arise from
loop processes at order $f^2/16\pi^2 = \alpha_f/4\pi$,
where $\alpha_f$ comes from some underlying flavor dynamics.
It need not have a high scale origin 
such as SUSY and/or GUTS \cite{RRR}.
We believe that the mass and mixing hierarchies,
with correlations as exemplified in   (\ref{eq:ansatz}),
cannot be just an accident.
\end{itemize}

\vskip 0.8cm
\noindent{\large\bf IV. Conclusion}
\vskip 0.3cm

We have covered two generic approaches to the 
quark mass-mixing problem.
Because of the scarcity of data, this is by far not 
the final word on this difficult but fascinating subject.
However, we also do not pretend that we know what happens at 
way beyond the weak scale, nor demand that physics far beyond the
weak scale is necessary to explain the flavor problem.

We think that the traditional Fritzsch path 
eventually runs into the difficulty of smallness of $V_{cb}$,
and remedies such as (\ref{eq:sob}) deviate drastically
from the old Weinberg Ansatz for the Cabibbo angle.
So, instead, we have emphasized and taken up the series 
expansion approach of Wolfenstein.
Let us summarize the salient features:

\begin{itemize}
\item In 1983, Wolfenstein parametrized $V_{ub} \sim \lambda^3$.
However, recent data give $V_{ub} \sim 0.003,\ V_{cb} \sim 0.04$ and
$m_t \sim 175$ GeV.
This leads to Eq. (\ref{eq:rho'}), that is
\[V_{ub} = A\lambda^4(\rho^\prime - i\eta^\prime)
\equiv B\lambda^4\, e^{-i\phi}\] where 
$\rho^\prime \equiv d\rho/d\lambda$ and $B \simeq 1.3$ {\it vs.} $A\simeq 0.8$.
So, from 1983 to 1993-1994, we have the change
\[ 
V_{ub}:\ \lambda^3 \longmapsto \lambda^4.
\]
This is the key starting point to revisit Wolfenstein's approach.
\item Turning to mass ratios, from Eq. (\ref{eq:mratio}), 
that is
\[
          {m_d\over m_s}  \sim  {m_s\over m_b} \sim \lambda^2
                                               \gg         
          {m_u\over m_c}  \sim  {m_c\over m_t} \sim \lambda^4.
\] \vskip -0.5cm
the mass ratios in the up-type sector are of higher order nature.
This in turn suggests that $V_{\rm KM}$ originates 
mostly from down-type sector, Eq. (\ref{eq:DL}),
while  $V_{ub}$ arises from up-type quarks, Eq. (\ref{eq:UL}).
Identifying $\delta \equiv A\lambda^2 \sim 0.04$ 
as the more appealing expansion parameter, we have\vskip -0.3cm
\[
\left\{ \begin{array}{l}
           D_L \equiv V_{\rm KM}(V_{ub} = 0) \ \ \ \ \Longleftarrow \ \ \ \
                    \hat M_d \cong \left[ \begin{array}{ccc}
                    0 & \delta^{3/2} &  0  \\
                    \delta^{3/2} & \delta &  \delta \\
                    0 & \delta & 1
                    \end{array} \right], \\
          U_L \equiv D_L\, V_{\rm KM}^\dagger \ \ \ \ \Longleftarrow \ \ \ \
\hat M_u \simeq
                    \left[ \begin{array}{ccc}
                    0 & 0 & -\delta^2\, e^{-i\phi} \\
                    0 & \delta^2 & 0 \\
                    -\delta^2\, e^{i\phi} & 0 & 1
                    \end{array} \right].
          \end{array}
\right.
\] \vskip -0.5cm
Wolfenstein's parameter $\lambda \sim \sqrt{\delta}$
is the more puzzling, arising empirically from 
the famous Weinberg texture zero of $(M_d)_{11} = 0$, Eq. (\ref{eq:Mad}),
which holds very well with 3 generation masses and mixings.
\item
Demanding the {\it least number of parameters},
which is related in spirit but not equivalent to finding
texture zeros, we suggest the following approximate relations
(Eqs. (\ref{eq:lambda}), (\ref{eq:Al2}) and (\ref{eq:Bl4})): \vskip -0.9cm
\begin{eqnarray}
   \lambda \ \,&=& \sqrt{m_d\over m_s + m_d}  = \vert V_{us}\vert,
            \nonumber\\
   A\lambda^2 &=& \ \ \, {m_s + m_d\over m_b}  = \vert V_{cb}\vert,
            \nonumber \\
   B\lambda^4 &=& \ \ \ \ \sqrt{m_u\over m_t} \ \ \,\,\, = \vert V_{ub}\vert,
              \ \ \ \ \ \ \ \ \ \ ({u\over c} = {c\over t} = B\lambda^4),  \nonumber
\end{eqnarray} \vskip -0.4cm
by idealizing the input equations 
(\ref{eq:rho'})--(\ref{eq:sob}).
\item Note that $A \sim 1 \sim B$, but
the smallness of $m_b/m_t$ is not explained.
Since  $m_b/m_t = (m_b/m_c)(m_c/m_t) 
\sim (m_b/m_c)\lambda^4 \sim A\lambda^3$,
perhaps both $A$ and the odd power $\lambda$
expansion are related to $m_b$ generation from a heavy top.
\item The placement of $CP$ phase in $(M_u)_{13}$ may 
be rooted in dynamics, since it naturally
{\it couples the smallness of $CP$ violation effects to the existence and
heaviness of the top quark}.
\item Since this ``numerology" seems to work well with
present, low energy data,
it suggests {\it WEAK SCALE ORIGIN!}

\end{itemize}

\vglue.3in
\noindent
{\bf Acknowledgments.}
The works reported here were done in collaboration with
Xiao-Gang He and Gwo-Guang Wong, respectively.





\begin{thebibliography}{99}
%
\bibitem{Fritzsch} H. Fritzsch, Phys. Lett. {\bf 70B}, 436 (1977).
%
\bibitem{HeHou} X. G. He and W. S. Hou, Phys. Rev. {\bf D 41}, 1517 (1990).
%
\bibitem{Wolf} L. Wolfensten, Phys. Rev. Lett. {\bf 51}, 1945 (1983).
%
\bibitem{HW} W. S. Hou and G.G . Wong, Phys. Rev. {\bf D 52}, 5269 (1995).
%
\bibitem{DHR} S. Dimopoulos, L. J. Hall, and S. Raby,
 Phys. Rev. Lett. {\bf 68}, 1984 (1991).
%
\bibitem{GSTCM} R. Gatto, G. Sartori, and M. Tonin,
 Phys. Lett. {\bf 28B}, 128 (1968); 
N. Cabibbo and L. Maiani, {\it ibid.} {\bf 28B}, 131 (1968).
%
\bibitem{Wein} S. Weinberg, Transactions of 
the New York Academy of Sciences, Series II, Vol. 38, 185 (1977).
%
\bibitem{KM} M. Kobayashi and T. Maskawa, Prog. Theor. Phys. {\bf 49},
652 (1973).
%
\bibitem{GJ} H. Georgi and C. Jarlskog,
Phys. Lett. {\bf 86B}, 297 (1979).
%
\bibitem{PDG} Review of Particle Properties,
Phys. Rev. {\bf D 50}, 1173 (1994).
%
\bibitem{sob} K. S. Babu and R. N. Mohapatra, Phys. Rev. Lett. 64, 2747 (1990).
%
\bibitem{Ng} D. Ng and Y. J. Ng, Mod. Phys. Lett. {\bf A6}, 2243 (1991).
%
\bibitem{Ma} E. Ma, Phys. Rev. {\bf D 43}, R2761 (1991).
%
\bibitem{Giudice} G. F. Giudice, Mod. Phys. Lett. {\bf A7}, 2429 (1992).
%
\bibitem{RRR} P. Ramond, R. G. Roberts, G. G. Ross, Nucl. Phys. {\bf B 406}, 19 (1993).
%
\bibitem{radm} For discussion of the lepton sector, 
see G. G. Wong and W. S. Hou, Phys. Rev. {\bf D 50}, R2962 (1994).
We ventured into quark mass--mixing numerology as an effort to
extend from leptons to quarks.
%
\end{thebibliography}
\end{document}